\documentclass[11pt,twoside]{article}

\usepackage{asp2006}
\usepackage{graphicx,amssymb}

\newcommand{\hi}{\mbox{H{\sc i}}}

\newcommand{\ha}{H$\alpha$}
\newcommand{\fantomm}{{\texttt {\textsc{FaNTOmM}}}}
\newcommand{\cigale}{{\texttt {\textsc{Cigale}}}}
\newcommand{\kmps}{km s$^{-1}$}

\begin{document}
\title{Dark matter in low mass surface density galaxies}   
\author{Laurent Chemin}\affil{Observatoire de Paris, section Meudon, GEPI, CNRS-UMR 8111 \& Universit\'e Paris 7, 5 Pl. Janssen, 92195 Meudon, France} 
\author{Claude Carignan}\affil{Laboratoire d'Astrophysique Exp\'erimentale, D\'epartement de physique,
Universit\'e de Montr\'eal, Montr\'eal, Qu\'ebec, Canada}    
\author{Philippe Amram}\affil{Observatoire Astronomique Marseille Provence, LAM, UMR 6110, 2 Pl. Le Verrier, 13248 Marseille, France}

\begin{abstract}
Low mass surface density spiral and irregular galaxies like low surface brightness (LSB) and
dwarf galaxies are unique laboratories to study the dynamical properties of Dark Matter (DM) halos
because their mass is generally dominated by DM at all galactocentric radii. 
We present results from the largest sample ever assembled of high resolution \ha\ velocity fields of LSB 
and dwarf galaxies in order to study their mass distributions.
\end{abstract}

\vspace*{-0.5cm}
\section{Observations}
Observations of 39 galaxies have been obtained at the OmM 1.6-m, OHP 1.93-m, CFHT and ESO 3.6-m telescopes
using Fabry-Perot (FP) interferometry with the \fantomm\  and \cigale\ instruments \citep{her03}, allowing typical resolving powers at \ha\ 
 between 8000-25000, spectral samplings between 7-16 \kmps, spatial samplings $\lesssim 1.5''$ and total scanning times of $\sim$ 3 hours per
galaxy. The sample contains LSB and dwarf galaxies chosen in previous studies from \cite{imp96}, \cite{mga01}, \cite{dbb02}, 
\cite{swa03} or \cite{mon03}. Galaxies for which DM halo shapes strongly disagree with the so-called ``universal" 
cosmological halo \citep{nav97} have been observed. Provisional results of the survey have been presented in \citet{che04} and \citet{gar05}. 
An adaptive binning is applied to the Fabry-Perot \ha\ cubes in order to improve the signal detection in low signal to noise regions,
 as described in \citet{che06}.  
 \begin{figure}
 \centering
\includegraphics[height=4.5cm]{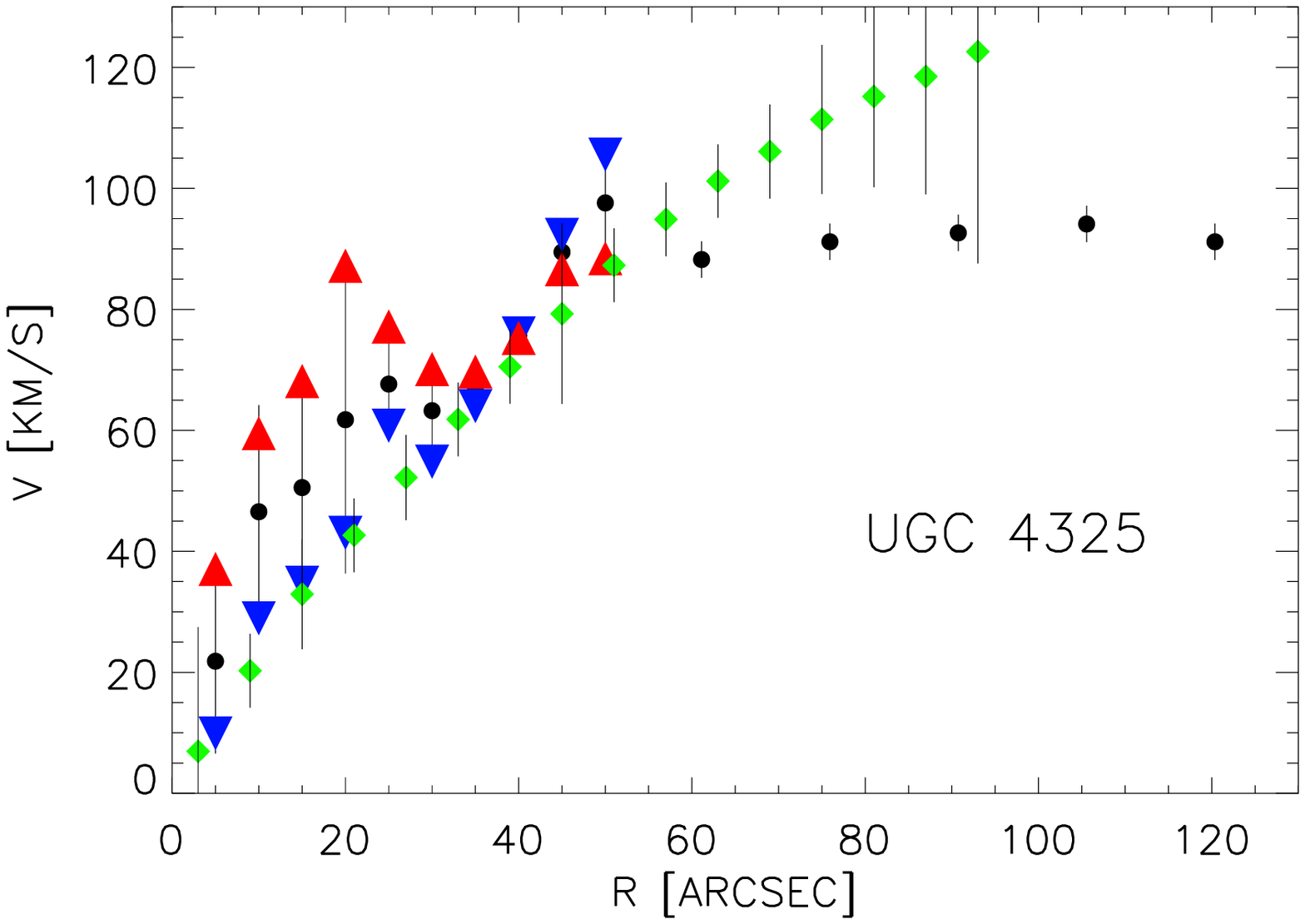}\includegraphics[height=4.5cm]{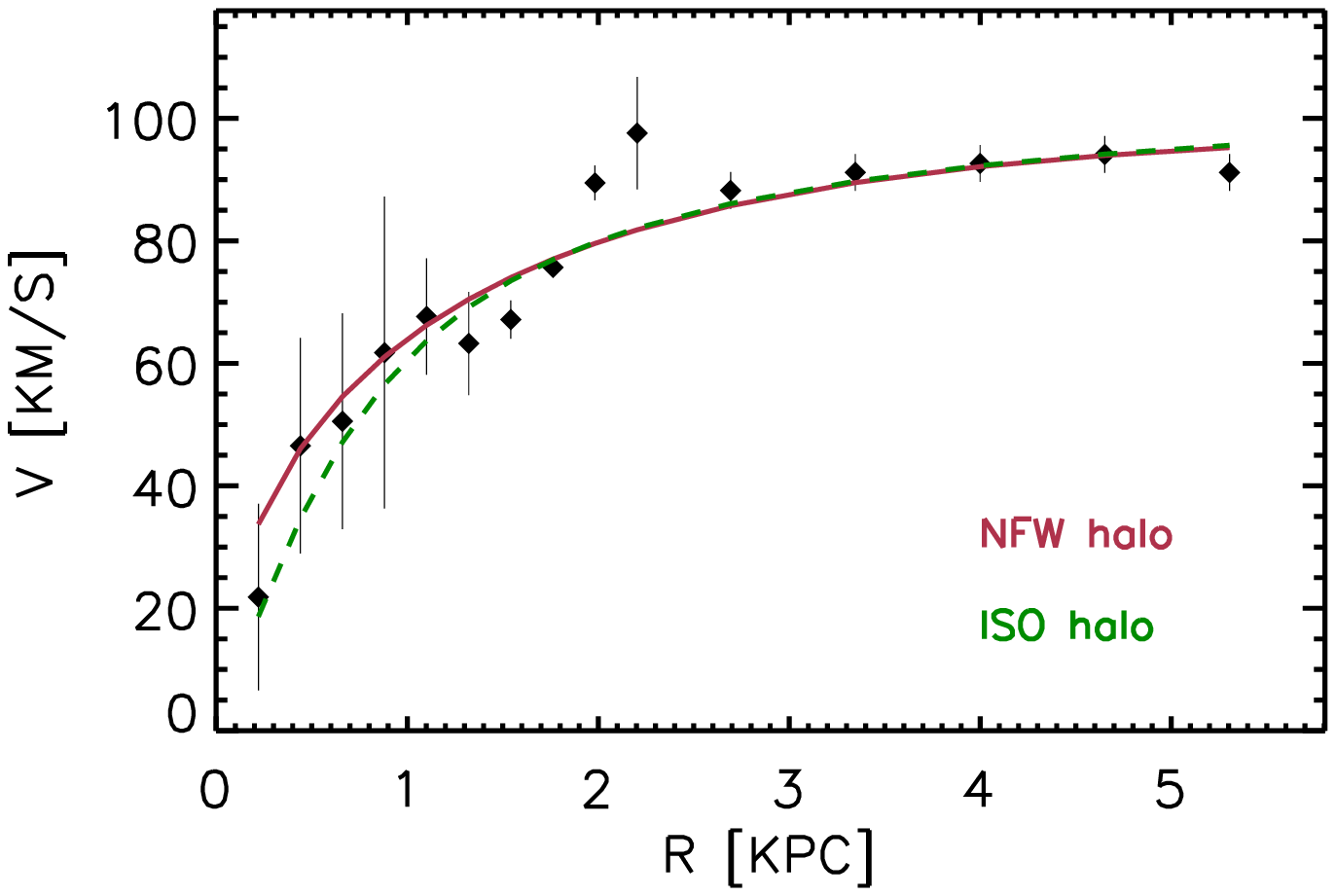}
\label{rc}
\caption{(Left panel) Observed rotation curve of UGC 4325. Blue downward (upward red) triangles are for the approaching (receding resp.) half of the disc, 
black circles for both sides fitted simultaneously and green squares are from slit spectroscopy data \citep{dbb02}. For $R > 50''$, black symbols are 
\hi\ measurements \citep{swa03}. (Right panel) Mass models using the pseudo-isothermal sphere (ISO, green dashed curve) and 
the Navarro-Frenk-White formalism (NFW, magenta curve).}
\end{figure}

\vspace*{-0.5cm}
\section{Results}
An example is illustrated in Fig.~\ref{rc} with the galaxy UGC 4325, which is supposed to exhibit a core-dominated halo according
to \citet{dbb02}. An hybrid rotation curve (RC) is derived from our \ha\ velocity field and \hi\ observations from \citet{swa03}.
The Figure also shows the  slit spectroscopy rotation curve from \citet{dbb02}. 
The FP velocities are larger than the slit spectroscopy ones in the innermost regions of the galaxy. 
A significant asymmetry is detected between the approaching and receding halves of the disc.   
The ionized gas motions are thus non-axisymmetric in these regions. The FP and slit spectroscopy \ha\ rotation curves are in good agreement 
in regions where the axisymmetry occurs ($\sim 30'' < R < 50''$).

Mass models of the rotation curve are shown in Fig.~\ref{rc} for the minimum stellar disc hypothesis. 
Two halo models are fitted, a pseudo-isothermal sphere and the Navarro-Frenk-White formalism. 
Because of the asymmetry detected in the gas motions, the FP error bars are large in this region of the galaxy. As a result, 
the halo fittings give equivalent results.   
This example illustrates that one clearly needs to model the gas motions taking into account perturbations of the potential 
if one properly wants to discriminate the dark matter halo shape within such low surface density galaxies.
These results will be presented in a forthcoming article, with the dynamical analysis of the whole sample.

\end{document}